\documentclass[12pt,a4paper,final]{iopart}

\usepackage[utf8]{inputenc}
\usepackage{enumitem,amssymb}
\usepackage[table,xcdraw]{xcolor}
\usepackage{graphicx,caption}
\usepackage[numbers,sort&compress]{natbib}
\usepackage{iopams}

\usepackage[section]{placeins}
\expandafter\let\csname equation*\endcsname=\relax 
\expandafter\let\csname endequation*\endcsname=\relax 

\usepackage{amsmath}

\usepackage{geometry}
\usepackage{graphicx}
\usepackage{subcaption}
\usepackage{hyperref}

\begin{document}

\title{{\bf{Quantum transport efficiency in noisy random-removal and small-world networks}}
}
\author{Arzu Kurt}
\address{Department of Physics, Bolu Abant \.{I}zzet Baysal University, 14030-Bolu, T\"{u}rkiye}
\ead{arzukurt@ibu.edu.tr}

\author{Matteo A. C. Rossi}
\address{QTF Centre of Excellence, Center for Quantum Engineering,
Department of Applied Physics, Aalto University School of Science, FIN-00076 Aalto, Finland}
\address{Algorithmiq Ltd, Kanavakatu 3C, FI-00160 Helsinki, Finland} 
\ead{matteo.rossi@aalto.fi }

\author[cor1]{Jyrki Piilo}
\address{Department of Physics and Astronomy, University of Turku, FI-20014 Turun Yliopisto, Finland}
\ead{jyrki.piilo@utu.fi}

\begin{abstract}
We report the results of an in-depth study of the role of graph topology on quantum transport efficiency in random removal and Watts-Strogatz networks. By using four different environmental models -- noiseless, driven by classical random telegraph noise (RTN), thermal quantum bath, and bath+RTN -- we compare the role of the environment and of the change in network topology in determining the quantum transport efficiency. We find that small and specific changes in network topology is more effective in causing large change in efficiency compared to that achievable by environmental manipulations for both network classes. Furthermore, we have 
found that noise dependence of transport efficiency in Watts-Strogatz networks can be categorized into six classes. 
In general, our results highlight the interplay that network topology and environment models play in quantum transport, and pave the way for transport studies for networks of increasing size and complexity -- when going beyond so far often used few-site transport systems.
\end{abstract}
 
\maketitle

\section{Introduction}
Understanding and controlling the efficiency of energy and charge transport in quantum systems is crucial for furthering the developments in quantum technologies and basic energy sciences. The role of the topology of the interaction network and its interplay with the detrimental environmental effects -- that influence the quantumness and the efficiency of the transport -- have been the subject of a large number of previous studies~\cite{Kim2003,Kay2006,Mulken2007,Bandyopadhyay2007,Tsomokos2008,Plenio2008,Rebentrost2009,Caruso2009,Bose2009,Scholak2009,Chin2010,Anishchenko2011,Mostarda2013,Mohseni2013,Novo2015,Walschaers2015,Baghbanzadeh2016,Bar2016,Celardo2016,Bassereh2017,Sierant2017,Rossi2017a,Abanin2017,Deng2018,Trautmann2018,Uchiyama2018,Harush2018,Mace2019,Nosob2019,Maier2019,Benedetti2019,Bueno2020,Mukai2020,Yalouz2020, Harush2020,Tian2020,Peng2021,Quiroz2021,Chisholm2021,cavazzoni2022perturbed}. Their results feature many surprising and counter-intuitive results, such as the enhancement of transport by environmental noise~\cite{Plenio2008,Rebentrost2009,Chin2010} or 
radical increase of efficiency with removal of a single edge in a completely connected network (CCN)~\cite{Anishchenko2011,Novo2015}. 

Many different mechanisms have been advanced to account for the efficiency of transport on quantum networks in excitation and state transfer contexts. Studies on basic interaction networks, such as linear, star or completely connected graphs -- without any external disturbance -- could be carried out analytically and provide basic understanding on the role of interference in the transportation of an excitation or a state from a source node to the sink node. References \cite{Caruso2009,Anishchenko2011,Novo2015,Bassereh2017} have shown that the efficiency of excitation transport between any two nodes in a CCN of size $N$ is inversely proportional to the number of nodes and could be increased to one by simply deleting the interaction between the source and the sink nodes. Using similar arguments, Bose et.~al.~\cite{Bose2009} have shown that the single link removal also allows perfect state transfer between the two nodes. The mechanism behind this counter-intuitive result is the elimination of destructive interference by removal of the single interaction~\cite{Caruso2009,Bose2009,Anishchenko2011,Novo2015,Bassereh2017,cavazzoni2022perturbed}. The problem could also be addressed by using graph spectral theoretical concepts in terms of the eigenvalues and eigenvectors of the adjacency matrix $A$ of the network. The eigensystem of $A$ for CCN of size $N$ is easy to construct: it has only two distinct eigenvalues of magnitude $N-1$ and $-1$ [$(N-1)$-fold degenerate]. References \cite{Bueno2020,Mukai2020} claim that null-eigenvalue localization is the mechanism responsible for the increased efficiency. A different approach where one tries to optimize the transport efficiency over a network with different edge weights has been undertaken by Scholak et.~al.~\cite{Scholak2009}.  Mostarda et.~al.~\cite{Mostarda2013} and 
Walschaers et.~al.~\cite{Walschaers2015} have found that the existence of centrosymmetry in the interaction matrix and  dominant doublet energy structure are required for robust high-efficiency transport. 

One of the important concepts, that has arisen in the efforts to explain the high efficiency excitation transport in photosynthetic systems, is the environment-assisted quantum transport (ENAQT), which refers to the enhancement of efficiency by interaction of the system with its environment~\cite{Plenio2008,Rebentrost2009,Chin2010}. Dephasing-induced delocalization~\cite{Rebentrost2009,Chin2010,Manzano2013}, line-broadening~\cite{Caruso2009}, super-transfer and funneling~\cite{Baghbanzadeh2016}, and super-radiance~\cite{Nesterov2013,Berman2015} are some of the mechanisms proposed for the existence of ENAQT. Furthermore, Zerah-Harush and Dubi~\cite{Harush2018,Harush2020} have shown that the dephasing-enabled density gradient and mixing of the system eigenstates are crucial for ENAQT. Recently, Chavez et.~al.~\cite{Chavez2021} have shown that more complicated noise dependence of transport efficiency might be observed in the nearest neighbor interacting system when additional long-range hopping is added. Studies on the noise dependence of the quantum transport efficiency phenomenon in interaction networks have considered mostly static noise~\cite{Tsomokos2008,Chin2010,Mohseni2013,Walschaers2015,Bar2016,Celardo2016,Baghbanzadeh2016,Sierant2017,Abanin2017,Deng2018,Trautmann2018,Maier2019,Mace2019,Nosob2019,Yalouz2020, Harush2020,Quiroz2021} or dephasing noise~\cite{Plenio2008,Caruso2009,Scholak2009,Chin2010,Mostarda2013,Mohseni2013,Bassereh2017,Trautmann2018,Harush2018}. The effect of dynamical noise on the transport efficiency has been the subject of relatively fewer studies~\cite{Uchiyama2018,Maier2019,Peng2021,Benedetti2016,Giusteri2016}. Furthermore, for some physical systems, such as photosynthetic energy transport apparatus, treating the environment as a single thermal bath, or approximating it as a classical Gaussian noise, is thought to be inadequate~\cite{Jan2018}. This leads to a crucial question: Is it possible to explore an active role that a geneneric environment -- consisting of both dynamical noise and quantum thermal baths --  has on quantum transport efficiency within complex networks?

In the present study, we investigate the excitation transport on random removal (Fig.~\ref{fig:scheme}(b)) and small-world networks (Fig.~\ref{fig:scheme}(c)) which are in contact with a thermal bath and driven by classical noise. Our purpose is to elucidate the effect of interplay between the topology, the environment, and the external driving on the transfer efficiency. Random removal networks are obtained by randomly deleting $n_R$ edges from a completely connected network of $N$ vertices. We find that removing the single link between the source and the sink sites, or removing any perfect matching of the graph~\cite{Bose2009}, leads to very high transport efficiency. 
For the Watts-Strogatz small-world network the highest transport efficiency is achieved when the range of interactions extends to $k=(N/2-1)$th neighbors, where $N$ is the number of sites of the system. We have also investigated the different types of noise response one could observe in Watts-Strogatz networks. It is found that while the efficiency of some realizations of these networks decreases monotonically with the strength of the noise, one could also observe noise-degraded, noise-enhanced (NET) and noise-independent transport (NIT) regimes on other realizations. This indicates, e.g., that the NET-NIT regimes of Ref.~\cite{Chavez2021} might be widespread. In general, our results for both the random removal and Watts-Strogatz networks indicate that a small and specific change in the network topology can lead to a very strong influence on the transport efficiency.

The paper is organized as follows. In Section \ref{sec:model}, we introduce the model on the graph, review the adopted master equation in the variational polaron frame, and introduce in detail the two types of networks we study. In Section \ref{sec:results}, we present the central results for the transport efficiency of these networks using different environmental models. Finally, we conclude and summarize in
Sec.~\ref{sec:conc}.

\section{Model}
\label{sec:model}

\begin{figure}[t]
    \centering
    \includegraphics[width=\textwidth]{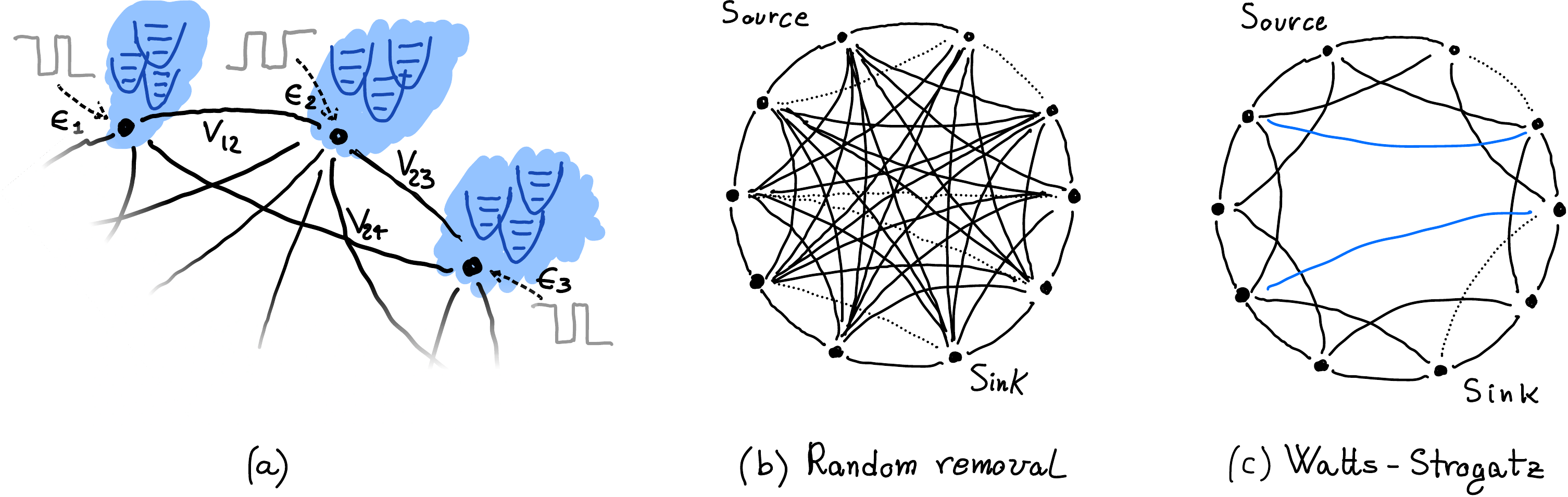}
    \caption{\textbf{Schematic representation of the model and networks.} (a): a depiction of the open quantum dynamics model, which consists of independent thermal baths of harmonic oscillators attached to each node, plus a perturbation to the energy of the node given by random telegraph noise (RTN). The noise is anti-correlated between neighboring sites. (b-c): the random networks considered in this work, with $N = 10$ nodes. In the random-removal model (b), we start from a complete graph and remove $n$ links, picked randomly (dotted lines). In the Watts-Strogatz model, (c), each node is connected to $2\,k$ neighbors (here $k = 2$). Each edge is then rewired with probability $p$: a node is replaced with another one, picked randomly, avoiding self-loops and edge duplication (the rewired edges are shown with a dotted line (initial target) and a blue line (new target).}
    \label{fig:scheme}
\end{figure}

We will first describe the network topology. Let us consider a graph $G(V,E)$ which is a collection of non-empty sets of vertices (nodes)  $V$ and edges (links) $E$. The adjacency matrix $A$ of $G$ illustrates the topology of the network by describing which nodes are connected by edges:
\begin{eqnarray}
\left[A(G)\right]_{ij}=\left\{
\begin{array}{ll}1& \textrm{If $i\neq j$ and $(i,j)\in E(G)$}\\
0&\textrm{otherwise}.
\end{array}\right.
\end{eqnarray}

The aim of the current study is to investigate the transport efficiency in random removal and Watts-Strogatz networks driven by classical noise, and also in combination with a thermal environment. The interest lies in the question of how network topology is related to transport efficiency within these noise bath models. Random removal networks are obtained from the CCN with $N$ nodes by randomly removing $n_R$ out of the total $N(N+1)/2$ edges (cf. Fig.~\ref{fig:scheme}(b)). There exists an interesting and well-known analytical result for the transport behavior of such a network~\cite{Caruso2009,Novo2015}; while the efficiency in CCN is $1/(N-1)$, removing only the edge between the source and the sink nodes increases it to 1, regardless of the number of nodes in the network. In the current study, however, we will go beyond this observation and scheme. The second type of network that we study is the Watts-Strogatz graph~\cite{Watts1998}. This was the first network model that successfully explains some important features of real-world or so-called small-world networks, i.e. the average path length growing logarithmically with the network dimension. Watts-Strogatz networks are characterized by three parameters: the number of nodes ($N$), the number of nearest neighbors on either side of a node included in edge connections ($k$) and the edge rewiring probability $p$ (cf. Fig~\ref{fig:scheme}(c)). When $p=0$, the network is a circular graph of $N$ nodes of degree 2 when $k=1$, and transforms into a completely connected network at $k=N/2$ (even $N$) or $(N-1)/2$ ($N$ odd). For a strict definition, when $N$ is even, $p=0$, and $k=N/2$, the Watts-Strogatz network would have double edges for the farthest neighbors, which disqualifies it as completely connected network. On the other hand, at $p=1$, all the connections of the network are rewired randomly and one obtains an Erdős–Rényi (ER) random graph~\cite{Mason2012}.

To formally describe the quantum transport in such a network, one should derive a master equation that defines how site populations change with time. In the present study, we will consider a multi-site spin-boson model whose interaction geometry is described by different realizations of two classes of widely studied geometries; namely, random removal and Watts-Strogatz networks.  Based on the variational polaron master equation for the multi-site spin-boson model derived by Pollock et.~al.~\cite{Pollock2013}, we have recently adopted this model by adding classical noise to the site energies~\cite{Kurt2020}. We showed that the adopted master equation is valid in the wide range of system-bath coupling strengths, which allows us to have the time-dependent site energies in the context of a multisite spin-boson model. Here we summarize the model while more technical details can be found in Ref.~\cite{Kurt2020}. A schematic visualization of the model is presented in Fig.~\ref{fig:scheme}(a).

We consider a noisy multi-site spin-boson model whose site energies are subject to a random telegraph noise (RTN) and whose total Hamiltonian is given by 
\begin{equation}\label{eq:ham}
    H=H_S(t)+H_B+H_I,
\end{equation}
where $H_S(t)$ is the Hamiltonian for the system, $H_B$ is the Hamiltonian for the quantum thermal bath and $H_I$ is the interaction between the two:
\begin{eqnarray}
H_S(t)&=&\sum_{n}\epsilon_n(t)|n\rangle\langle n|+\sum_{n\neq m}V_{nm}|n\rangle\langle m|,\label{eq:ham_s}\\
H_B&=&\sum_{n,k}\omega_{n,k}\,b_{n,k}^{\dagger}b_{n,k},\nonumber\\
H_I&=&\sum_{n,k}|n\rangle\langle n|(g_{n,k}b_{n,k}^{\dagger}+g_{n,k}^{*}b_{n,k}),\nonumber
\end{eqnarray}
where $V_{nm}$ is the electronic coupling between $n$th and $m$th nodes, $b_{n,k}(b_{n,k}^{\dagger})$ is the annihilation (creation) operator for
the $k$th oscillator mode of the $n$th node whose state is described by $|n\rangle$, while $g_{n,k}$
is the strength of interaction between the mode at $n$th node and the $k$th oscillator of its environment. We suppose that the site energy of each node is modulated by an RTN, i.e.,
$\epsilon_{n}(t)=\epsilon_{n0}+\Omega_n\alpha_n(t)$, where $\epsilon_{n0}$ is the static site energy of $n$th node and $\Omega_{n}$ is the noise amplitude at
$n$th node. The RTN is a stochastic process that flips between two possible values $\alpha = \pm 1$ with a given rate $\nu$. It 
is also described by two parameters: a zero mean ($ \langle\alpha_n(t)\rangle=0$) and an exponentially decaying auto-correlation functions ($\langle\alpha_n(t)\alpha_n(t')\rangle=e^{-\nu|t-t'|}$). Here, the correlation time of the noise is $\tau_{c}=1/\nu$.

In the present study, we treat the bosonic environment of the system as a set of independent harmonic oscillators. The spectral density, $J_n(\omega)=\sum_{k}|g_{n,k}|^{2}\delta(\omega-\omega_k)$, contains all the information to describe the system-bath interaction as well as the spectral properties of the environment. Reorganization energy, $E^{r}_n=\int_{0}^{\infty}d\omega\,J_n(\omega)/\omega$, is a measure of the strength of system-environment coupling. For the current study, we will use a specially structured spectral density $J_{com}(\omega)$ that was developed to account for the complex environmental effects which might be relevant for the bath-enhanced transport. It is sum of an overdamped and broad background and a discrete vibrational mode that itself interacts with an Ohmic environment with cut-off frequency $\Lambda$~\cite{Roden2012,Kell2013}:
\begin{eqnarray}
\label{eq:spectJcom}
    J_{com}(\omega)&=&J_{bg}(\omega)+J_{vib}(\omega),\\
    J_{bg}(\omega)&=&\sqrt{\frac{\pi}{2}}\frac{S \omega}{\sigma}\exp\left[-\frac{1}{2}\left(\frac{\log{[\omega/\omega_ c]}}{\sigma}\right)^2\right]\label{eq:spectJbg}\nonumber,\\
    J_{vib}(\omega)&=&X\omega^{2}\frac{J_{ohm}(\omega)}{(\omega-g(\omega))^{2}+J_{ohm}(\omega)^{2}}\label{eq:spectJvib}\nonumber,\\
    J_{ohm}(\omega)&=&\xi\,\omega\,e^{-\omega/\Lambda},\quad g(\omega)=\zeta-\xi\,\frac{\Lambda}{\pi}+\frac{1}{\pi}J_{ohm}(\omega)\mathrm{Ei}[\omega/\Lambda]\nonumber,
\end{eqnarray}
where $S$ and $X$ are measures of the magnitude of the background $J_{bg}(\omega)$ and vibrational $J_{vib}(\omega)$ spectral functions, respectively. Here, the cutoff and dispersion of $J_{bg}(\omega)$ are denoted by $\omega_c$ and $\sigma$. The more complicated damped discrete vibrational mode has $\xi$ and $\Lambda$ in the role of damping factors and $\zeta$ determines the position of the discrete mode. $\mathrm{Ei}[x]$ in the last line is the exponential integral function. Note that each node is assumed to have the same thermal environment with spectral function parameters: Background cutoff frequency $\omega_c=1$, standard deviation $\sigma=0.7$, peak amplitude factor $S=0.06\;x_f$, Huang-Rhys factor $X=0.025\;x_f$, damping factor $\xi=0.3$, cutoff frequency for Ohmic bath $\Lambda=5$ and center frequency of vibrational mode $\zeta=5$. $\omega_c$, $\Lambda$, and $\zeta$ have energy units, while the other parameters are dimensionless. $x_f$ is a scaling factor that determines the strength of the system-bath interaction.

Usually, the Redfield master equation is used when the system-bath interaction is weak~\cite{Redfiled1957,Breuer2002}, while the full polaron master equation \cite{Jackson1983} is useful when it is large. On the other hand, the reliable regime of energy transport efficiency in realistic systems such as light-harvesting Fenna-Matthews-Olsen (FMO) complexes would be the intermediate one~\cite{Grondelle2006}. The reason why we choose the intermediate regime in the present study is the fact that a large number of widely used master equations for the weak and strong coupling regimes exist, while the intermediate coupling regime has not been studied as widely as those two regimes. Hence, to describe the dynamics of the reduced density matrix of the current system, we use the variational master equation including the time-dependent function in the site energies~\cite{Pollock2013,Kurt2020}. 

By transferring the Hamiltonian in Eq.~(\ref{eq:ham}) into the variational polaron frame, the site energy $\epsilon_{n}(t)$ (the electronic coupling $V_{nm}$) is renormalized by $R_{n}$($\mathcal{B}_{n}$). For simplicity, we do not present the form of the Hamiltonian in the variational frame (see Ref.\cite{Pollock2013,Kurt2020}) and express the master equation for the reduced density matrix $\tilde{\rho}_{S}(t)=\mathrm{Tr}_{E}[\tilde{\rho}(t)]$ in the Sch\"{o}dinger picture:

\begin{align}
\label{eq:master}
    \frac{\partial \tilde{\rho}_{S}(t)}{\partial t} = & -i\left[\tilde{H}_{S}(t),\tilde{\rho}_{S}(t)\right]-i\left[\tilde{H}_{trap},\tilde{\rho}_{S}(t)\right]\notag\\
    &-\sum_{i,j=1}^{N^2}\int_{0}^{t}ds\langle E_{i}(s)E_{j}(0)\rangle\left(\{S_i\,S_j(s)\,\tilde{\rho}_{S}(t)-S_j(s)\tilde{\rho}_{S}(t)S_i\}+ hc\right).
\end{align}
Here, $\tilde{O}$ indicates that the operator is in the variational frame. It is important to note that the interaction Hamiltonian $\tilde{H}_{I}$ in the variational polaron frame is assumed to be $\tilde{H}_{I}=\sum_{i,j=1}^{N^2}S_i\otimes E_i$, where $N$ is the number of nodes, $S_{i}$ and $E_i$ are the $i$th system and bath operators, respectively. The term $\tilde{H}_{trap}=-i\,\kappa\,|s\rangle\langle s|$ is the anti-Hermitian trap Hamiltonian~\cite{Rebentrost2009,Fassioli2010,Montiel2014}, which is responsible for dissipation of the excitation from the sink node $|s\rangle$. $\langle E_{i}(s)E_{j}(0)\rangle$ are the bath correlation functions. $S_j(t)=U(t)S_j U^{\dagger}(t)$ is the $j$th time dependent system
operator in the interaction picture that is calculated as  
$U(t)=\mathcal{T}\exp{\left(-i\int_{0}^{t}H_{S}(t')dt'\right)}$ where
$\mathcal{T}$ is the time-ordering operator. Since the Hamiltonian of the system at different times does not commute due to the site energy fluctuations, the time evolution operator $U(t)$ cannot be obtained analytically. Details of the noise averaging of the master equation and $U(t)$ can be found in Ref.\cite{Kurt2020}.

\section{Results}
\label{sec:results}

All results presented below, except when stated otherwise, are for networks of ten vertices with zero site energy~i.e., $\epsilon_{n0}=0$ with $n=1,2,\dots, N$. The electronic coupling between the connected vertices is uniform with $V=2$, i.e., we assume that all types of networks are unweighted. When the thermal environment has been taken into account, the environment of each node is the same as that of the others with $k_BT=1$ and the system bath coupling is in the intermediate regime. The external RTN signal is assumed to act on the system collectively, which means that each node experiences the same noise sequence. Since such collective noise would have no effect on the dynamics, the sign of RTN at consecutive nodes is reversed, which would make the noise correlated (anticorrelated) for the even (odd) numbered neighbors (sort of "antiferromagnetic" like). The motivation for using anti-correlated noise amplitude is that a global disturbance, such as the conformational motion of the protein scaffold, might increase the site energy of some nodes while decreasing the energy of others, which makes it a natural external noise effect on system dynamics~\cite{Uchiyama2018, Huo2012, Bhattacharyya2013}. Calculations involving noise are carried out by using ensemble averaging, for which the solutions of Eq.~(\ref{eq:master}) are found for 100 RTN samples and averaged to obtain the noise-averaged density matrix. 

Our aim is to study how transport efficiency changes in the above-described multisite spin-boson systems. The transport efficiency $\eta$ basically describes the fraction of the source population that has been transferred to the target site within a given interval of time and the mathematical definition is the following~\cite{ Rebentrost2009,Uchiyama2018}:

\begin{equation}
    \eta=2\kappa\int_{0}^{t_{up}}\rho_{ss}(t)\,dt,
\end{equation} 
\noindent where $t_{up}$ is the upper time limit and $\kappa$ is the trapping rate at sink site $s$. $\eta$ is the probability of the excitation being trapped at the sink site in total time $t_{up}$. $\eta=1$ means the excitation is trapped at the sink site within $[0,t_{up}]$ time interval for sure. Various schemes to choose $\kappa$ and $t_{up}$ exist in the literature~\cite{ Rebentrost2009,Uchiyama2018,Kurt2020}. A larger $t_{up}$, in general, leads to a larger $\eta$ but it also increases the computational complexity of the problem when the thermal bath is included. In the present study, we use the values $\kappa=0.5$~\cite{Pelzer2013} and $t_{up}=\pi$. The source node is chosen as node number 1 and the sink node $s$ is chosen such that the graph distance between the sink and the source is the largest in the network.

\begin{figure}[t!]
\centering
 \begin{subfigure}{.4\textwidth}
 \includegraphics[width=\linewidth]{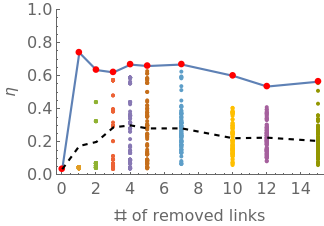}
 \caption{Noiseless} 
 \end{subfigure}
\centering
\begin{subfigure}{.4\textwidth}
 \includegraphics[width=\linewidth]{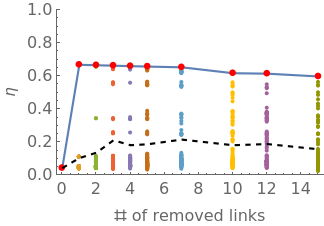}
 \caption{RTN-only}
\end{subfigure}
\centering
 \begin{subfigure}{.4\textwidth}
 \includegraphics[width=\linewidth]{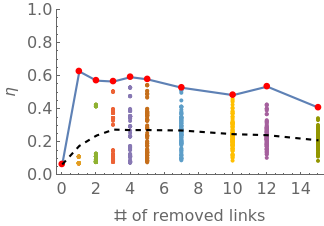}
 \caption{Bath-only} 
 \end{subfigure}
\centering
\begin{subfigure}{.4\textwidth}
 \includegraphics[width=\linewidth]{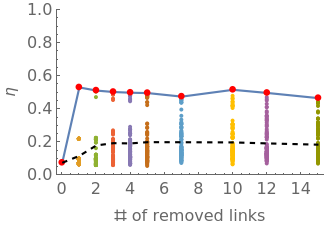}
 \caption{Bath+RTN}
\end{subfigure}
 \caption{Efficiency for the 10-site system as a function of number of removed links ($n_R$) for noiseless (a), RTN-only (b), bath-only (c), and bath+RTN (d) at the system parameters: $\Omega=10$, $\nu=1$, $V=2$, $k_BT=1$ and $x_f=2$. Different color dots in each graph indicate the number of edges removed from CCN. The straight blue line joins the maximum of efficiencies obtained for each $n_R$. Colored dots represent the efficiency of different network realizations for each $n_R$.  Black dashed lines in each sub-figure illustrate the mean efficiency. }
 \label{fig:random-removal-all-cases-eff}
\end{figure}

\subsection{Transport efficiency  and environmental models}

In this section, we study the transport efficiency with respect to the number of deleted links ($n_R$) and $k$th neighbor interaction for random removal and Watts-Strogatz networks, respectively. We consider four different environmental models: (i) noiseless (no noise and no bath), (ii) RTN-only, (iii)  bath only (nodes are in contact with the thermal bath only), and (iv) bath+RTN (a combination of thermal bath and classical noise acting on the system nodes).

We first present the transport efficiency versus the number of removed edges in the random removal networks in Figs.~\ref{fig:random-removal-all-cases-eff}(a)-(d) for the four noise configurations mentioned above.  The data displayed in the plots are obtained as follows: since the number of different ways to choose $n_R$ links to remove from a total of $N(N-1)/2$ links grows very fast with $n_R$, we randomly sample the space of possible link removals  when $n_R>2$, and calculate the efficiency of those sampled networks and display them as the colored points in Figs.~\ref{fig:random-removal-all-cases-eff}(a)-(d). The RTN used in Figs.~\ref{fig:random-removal-all-cases-eff} (b) and (d) has intermediate noise frequency ($\nu=1$) and high noise amplitude ($\Omega=10$). It is obvious that link removal, on average, increases efficiency independent of the environmental model, which was also reported in Refs.~\cite{Caruso2009,Novo2015,Anishchenko2011,Bassereh2017}. In all four cases, the highest efficiency is found when only the single link between the source and the sink nodes is removed as can be seen from Fig.~\ref{fig:random-removal-all-cases-eff}(a)-(d). We have found that a similarly high transport efficiency could be obtained by removing the "perfect matching," which is a disjoint set of edges that includes all the nodes in the network as expected~\cite{Bose2009}. These observations indicate that a small or strategic change in the network topology might affect the transport efficiency in a much stronger fashion than any environmental manipulations. The system-thermal environment interaction and the temperature ($k_{B}T=1$) in Figs.~\ref{fig:random-removal-all-cases-eff}(c) and (d) are assumed to be intermediate, so the effect of the bath on the efficiency is not significant. We have also done some calculations in the strong coupling regime for a small subset of networks, and the effect of bath in the strong interaction regime is comparable to that in the intermediate regime. We present the temperature and system-environment coupling dependence of the transport efficiency for the bath-only environment model in the Appendix. Comparing Figs.~\ref{fig:random-removal-all-cases-eff}(a) and (c) which display the efficiencies for the noiseless dynamics and the bath-only model, respectively, one can see that effect of the bath is to increase $\eta$ for the CCN ($n_R=0$) and decrease the maximum $\eta$ for almost all removals ($n_R>0$). The first observation might be attributed to the fact that including the environmental effects would lift the degeneracy in the network spectrum and inhibit the massive destructive interference in the CCN~\cite{Caruso2009}. If one considers only the maximum efficiency attainable for each number of removed links (straight lines in Figs.~\ref{fig:random-removal-all-cases-eff}(a)-(d)), including the thermal environment, the classical RTN or a combination of the two, generally, leads to a decrease regardless of the number of removed links. 

\begin{figure}[t!]
\centering
 \begin{subfigure}{.4\textwidth}
 \includegraphics[width=\linewidth]{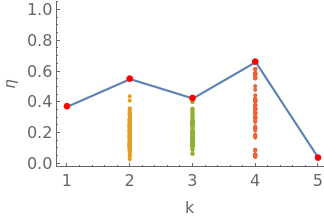}
 \caption{Noiseless} 
 \end{subfigure}
\centering
\begin{subfigure}{.4\textwidth}
 \includegraphics[width=\linewidth]{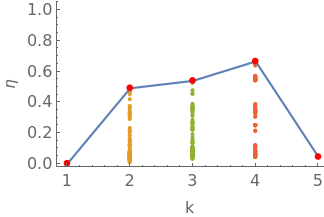}
 \caption{RTN-only}
\end{subfigure}
\centering
 \begin{subfigure}{.4\textwidth}
 \includegraphics[width=\linewidth]{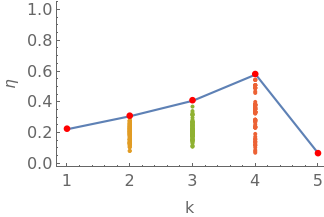}
 \caption{Bath-only} 
 \end{subfigure}
\centering
\begin{subfigure}{.4\textwidth}
 \includegraphics[width=\linewidth]{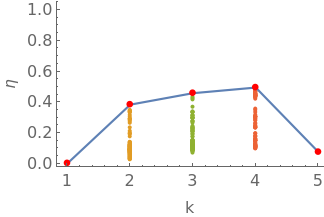}
 \caption{Bath+RTN}
\end{subfigure}
 \caption{Efficiency versus $k$ for noiseless (a), RTN-only (b), bath-only (c), and bath+RTN (d) for the Watts-Strogatz network at rewiring probability $p=0.75$. The parameters of the thermal environment and the classical noise are the same as those used in  Fig.~\ref{fig:random-removal-all-cases-eff}. The straight line joins the maximum of efficiencies obtained for each $k$ value. Colored dots represent efficiency of different network realizations for the given $k$.}
 \label{fig:watts_eff_vs_k_allcses_nSite_10}
\end{figure}

The second type of network we study is the Watts-Strogatz small-world network. The efficiencies as a function of $k$ (the nearest neighbors
included in edge connections) for different network realizations, with  $p=0.75$,  are displayed for noiseless, RTN-only, bath-only and bath+RTN combinations in Figs.~\ref{fig:watts_eff_vs_k_allcses_nSite_10}(a)-(d), respectively. The data of the plots in the figure for $k=2,3,$ and 4 are obtained by generating 100 different networks with re-wiring probability $p=0.75$ for each $k$ value. $k=1$ and $k=5$ in the plots correspond to circular and completely connected networks, respectively. Similarly, Fig.~\ref{fig:random-removal-all-cases-eff}, colored dots  illustrate the transport efficiency of different network realizations for each $k$. The most prominent finding from the results in Fig.~\ref{fig:watts_eff_vs_k_allcses_nSite_10} is that 
the maximum efficiency is obtained when the interactions up to the $k$th nearest neighbor are turned on where $k$ is one layer away from making the network completely connected, independent of the environmental model.  Although the results are reported only for ten nodes here, we have checked this finding for the noiseless and RTN-only models for $N$ up to 100 and found the same result. Here similar to the results for the random removal network in Fig.~\ref{fig:random-removal-all-cases-eff}, one can also see that the maximum  of efficiency is mainly determined by the change in topology rather than by the environmental conditions. One could deduce from a comparison of Fig.~\ref{fig:watts_eff_vs_k_allcses_nSite_10}(a)-(c) that the effect of the bath as well as the external noise on the $\eta$ in the circular and completely connected networks are opposite; they enhance the transport for the CCN and degrade it for the circular one. Another important observation from Fig.~\ref{fig:watts_eff_vs_k_allcses_nSite_10} is that the maximum efficiency increases uniformly with increasing $k$ up to and including $k=4$ for noisy (thermal, external, and combination of the two) networks and is always lower when the network is completely connected. Thus, the maximum of $\eta$ as a function of $k$ display a broad resonance structure. 

\begin{figure}[t!]
\centering
 \begin{subfigure}{.27\textwidth}
 \includegraphics[width=\linewidth]{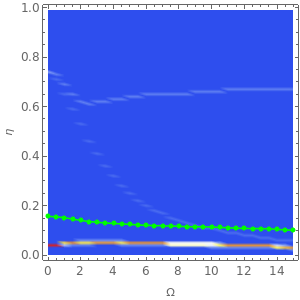}
 \caption{$n_R=1$} 
 \end{subfigure}
 \centering
 \begin{subfigure}{.27\textwidth}
 \includegraphics[width=\linewidth]{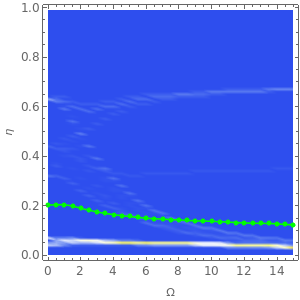}
 \caption{$n_R=2$} 
 \end{subfigure}
\centering
 \begin{subfigure}{.27\textwidth}
 \includegraphics[width=\linewidth]{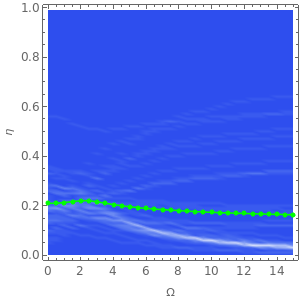}
 \caption{$n_R=15$} 
 \end{subfigure}
\centering
\begin{subfigure}{.27\textwidth}
 \includegraphics[width=\linewidth]{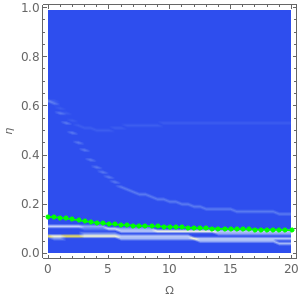}
 \caption{$n_R=1$}
\end{subfigure}
\centering
 \begin{subfigure}{.27\textwidth}
 \includegraphics[width=\linewidth]{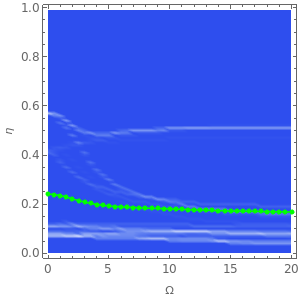}
 \caption{$n_R=2$} 
 \end{subfigure}
\centering
\begin{subfigure}{.27\textwidth}
 \includegraphics[width=\linewidth]{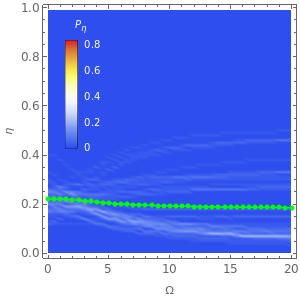}
 \caption{$n_R=15$}
\end{subfigure}
 \caption{Efficiency distribution as a function of noise amplitude $\Omega$ for a set of random removal networks obtained by removing $n_R=1,2,15$ links from the completely connected network for RTN-only model (first row) and bath+RTN model (second row) at the same system parameters used as in Fig.~\ref{fig:random-removal-all-cases-eff}. Green dotted lines for each sub-figure refer the efficiency averaged over different realizations of random removal networks.}
 \label{fig:noise-only-remove-randomly-eff-dist}
\end{figure}

\subsection{Efficiency distribution}
We now investigate the transport efficiency distribution over the network realizations for different values of the amplitude 
$\Omega$ and for the two different network groups which were introduced previously. 
The results for $\eta$ distributions with RTN frequency $\nu=1$ at intermediate temperature and system-bath coupling regime are shown in Figs.~\ref{fig:noise-only-remove-randomly-eff-dist}(a)-(f) for random removal network with $n_R=1,2,15$. The upper row corresponds to the RTN-only model and the lower row for the bath+RTN model.  The average over the network realizations for fixed $n_R$ and $\Omega$ is shown as green dots.
The results show that $\langle\eta\rangle$ decreases with increasing noise strength independent of the number of removed edges and the environmental model. One of the interesting observations from these plots is that the excitation transport in a subset of networks generated by removing $n_R$ links from CCN is enhanced by the external noise as can be deduced from the white lines in the plots that increase with the increasing noise strength $\Omega$. The enhancement is especially pronounced when the number of removed edges is high [Figs.~\ref{fig:noise-only-remove-randomly-eff-dist} (c) and (f)]. This finding is probably due to the fact that the probability of generating networks with unique eigensystems is higher when the number of links to be removed is high and the external noise increases efficiency in some of those networks.

\begin{figure}[t!]
\centering
 \begin{subfigure}{.4\textwidth}
 \includegraphics[width=\linewidth]{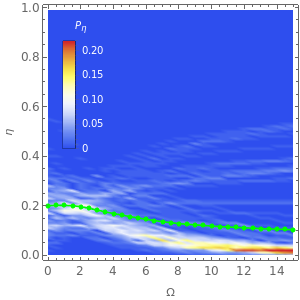}
 \caption{RTN-only} 
 \end{subfigure}
 \centering
\begin{subfigure}{.4\textwidth}
 \includegraphics[width=\linewidth]{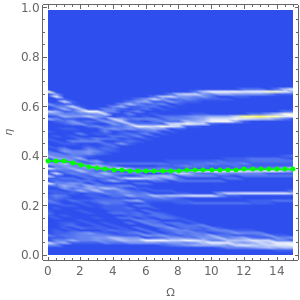}
 \caption{RTN-only}
\end{subfigure}
\centering
 \begin{subfigure}{.4\textwidth}
 \includegraphics[width=\linewidth]{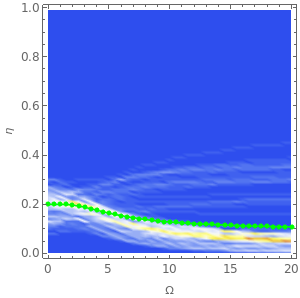}
 \caption{Bath+RTN} 
 \end{subfigure}
\centering
\begin{subfigure}{.4\textwidth}
 \includegraphics[width=\linewidth]{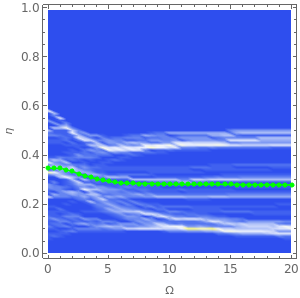}
 \caption{Bath+RTN}
\end{subfigure}
 \caption{Efficiency distribution as function of noise amplitude $\Omega$ for Watts-Strogatz network at different $k=2$ (a,c) and $k=4$ (b,d) for RTN-only (first row) and bath+RTN (second row) at $p=0.75$. The parameters of the thermal environment and the classical noise are the same as those used in  Fig.~\ref{fig:random-removal-all-cases-eff}. Green dotted lines for each sub-figure refer the efficiency averaged over different realizations of Watts-Strogatz networks.}
 \label{fig:watts_effDist_vs_Om_NO_NPB_nSite_10}
\end{figure}

Similar efficiency distribution plots for RTN-only (first row) and bath+RTN (second row) environment models are displayed in Fig.~\ref{fig:watts_effDist_vs_Om_NO_NPB_nSite_10}(a)-(d) for Watts-Strogatz networks with $k=2,4$ and rewiring probability $p=0.75$.  The efficiency averaged over all network realizations (green dotted lines in Fig.~\ref{fig:watts_effDist_vs_Om_NO_NPB_nSite_10}) is found to decrease monotonously with the noise amplitude for $k=4$ at small noise amplitude and remains independent of $\Omega$ as the noise strength is increased further while it shows a very small increase at low noise strength (Fig.~\ref{fig:watts_effDist_vs_Om_NO_NPB_nSite_10}(a)) and then decreases with increasing $\Omega$ for $k=2$ and RTN-only environmental model. Similar to the random removal networks discussed above, the transport efficiency on some of the small-world networks generated with $p=0.75$ is enhanced by external noise while $\eta$ on others is degraded by the same noise as can be deduced from the white streaks in all four subplots in the figure. The external noise dependence of $\eta$ distribution displays an interesting banded structure, especially for $k=4$ both for the RTN-only and bath+RTN environment models.

Figs.~\ref{fig:noise-only-remove-randomly-eff-dist} and \ref{fig:watts_effDist_vs_Om_NO_NPB_nSite_10} illustrate that the noise dependence of the transport efficiency could be categorized into a small number of classes. Although intuitively, one would expect the efficiency to decrease with increasing noise strength, many studies have indicated that there might be certain conditions under which $\eta$ displays a resonance structure with noise strength (ENAQT)~\cite{Plenio2008,Rebentrost2009, Caruso2009, Mohseni2013}. Besides, a recent study by Chavez et.~al.~observes that the static noise dependence of the transport efficiency for long-range hopping problems could be categorized as disorder-degrading, disorder-enhanced and disorder-independent transport regime~\cite{Chavez2021}. A natural question that arises is could there be  any other behavior of efficiency as function of the noise strength? To answer the question, we turn our attention to the classification of transport efficiency. We note that the findings presented in the rest of the paper are obtained for the RTN-only model, which does not take into account the quantum thermal bath.

\subsection{Classification of transport efficiency for Watts-Strogatz networks}
 
\begin{figure}[t!]
\centering
 \begin{subfigure}{.27\textwidth}
 \includegraphics[width=\linewidth]{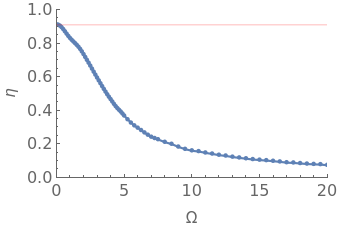}
 \caption{MD} 
 \end{subfigure}
 \centering
 \begin{subfigure}{.27\textwidth}
 \includegraphics[width=\linewidth]{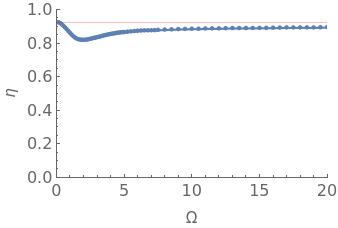}
 \caption{DI} 
 \end{subfigure}
 \centering
 \begin{subfigure}{.27\textwidth}
 \includegraphics[width=\linewidth]{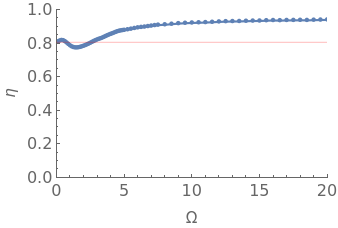}
 \caption{DI2} 
 \end{subfigure}
 \centering
 \begin{subfigure}{.27\textwidth}
 \includegraphics[width=\linewidth]{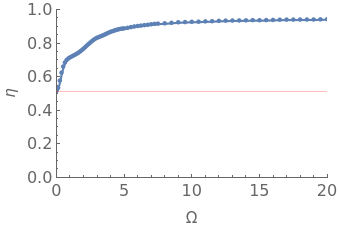}
 \caption{I} 
 \end{subfigure}
 \centering
 \begin{subfigure}{.27\textwidth}
 \includegraphics[width=\linewidth]{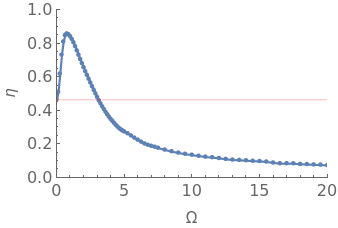}
 \caption{ID} 
 \end{subfigure}
 \centering
 \begin{subfigure}{.27\textwidth}
 \includegraphics[width=\linewidth]{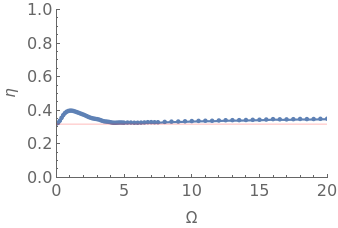}
 \caption{IDI} 
 \end{subfigure}
 \centering
 \caption{Different types of noise dependence of the transport efficiency observed for the Watts-Strogatz networks. (a) monotonic decay (MD), (b) first decreasing then increasing to a value lower than the noiseless transport (DI),
 (c) first decreasing then increasing to a value higher than the noiseless transport (DI2), (d)  increasing (I), (e) first increasing then decreasing (ID), (f) increasing-decreasing-increasing (IDI). The displayed plots are chosen as examples of the different classes from $N=16$ and $k=7$ and each is computed for a single network.}
 \label{fig:watts-class}
 \end{figure}

 We now investigate, in detail, the noise strength dependence of the transport efficiency in 16-node Watts-Strogatz networks with different $k$ and $p$ values under the influence of the RTN. For the study presented here, we have generated a total of 7200 different realizations of the network for each $k\in[1,7]$  at 10 different values of $p$ (in the range [0,1]) and investigated the noise amplitude dependence of the efficiency of each generated network for $\Omega\in[0,20]$. The resulting data sets were classified based on the behaviour of $\eta$ as function of $\Omega$. Surprisingly, we have found that one could classify $\Omega$-dependence of $\eta$ into six different classes as displayed in Fig.~\ref{fig:watts-class}. Although the displayed plots are for $N=16$ and $k=7$ with different $p$ values, the same types of noise dependence are observed for different $N$, $k$ and $p$ values, as well. The six classes could be described as: (i) Monotonic decay (MD) [Fig.~\ref{fig:watts-class}(a)], (ii,iii) first decreasing then increasing to a value lower (higher) than the noiseless transport DI (DI2) [Fig.~\ref{fig:watts-class}(b), (Fig.~\ref{fig:watts-class}(c)], (iv) increasing I [Fig.~\ref{fig:watts-class}(d)], (v) first increasing then decreasing ID [Fig.~\ref{fig:watts-class}(e)], and (vi) increasing-decreasing-increasing IDI [Fig.~\ref{fig:watts-class}(f)]. In MD, DI, and ID classes, at high noise intensity, the transport efficiency is lower than that for the noiseless dynamics (at $\Omega=0$ in Fig.~\ref{fig:watts-class}), 
while the opposite holds for the DI2, I and IDI classes. The typical behaviour of ENAQT corresponds to ID class in Fig.~\ref{fig:watts-class}(e) where one could observe that  transport is enhanced (degraded) by weak (strong) noise. Noise-degraded transport-NET-NIT type noise-efficiency dependency reported by Ref.~\cite{Chavez2021} is similar to Fig.~\ref{fig:watts-class}(b). Also, it should be noted that I, DI2 and IDI classes also display noise-enhanced transport but contrary to ENAQT, the high noise intensity does not degrade efficiency in these networks.

\begin{figure}[t!]
\centering
 \includegraphics[width=0.75\linewidth]{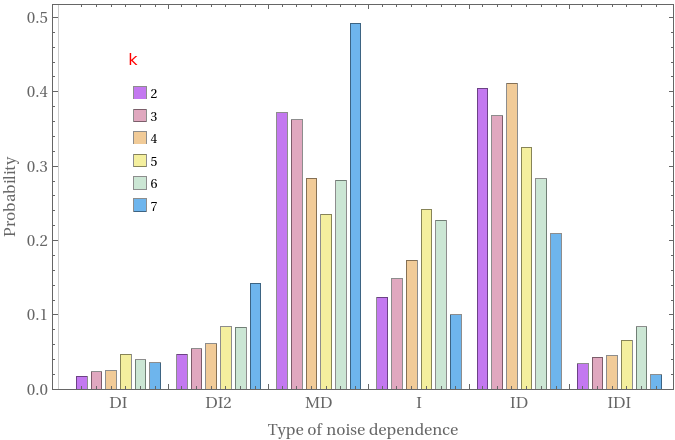}
 \caption{Distribution of different noise dependence types in the ensemble of Watts-Strogatz networks with $N=16$ for different $k$ values and with all values of $p\in[0,1]$ with step size 0.1.}
 \label{fig:watts-dist}
\end{figure}

Given a large number of realizations of the networks, we can also study the distribution of the realizations over the six different classes mentioned above; see Fig.~\ref{fig:watts-dist}, where we use $N=16$ and $k=2,\ldots,7$. It can be observed from the figure that the efficiency of noise-dependent transport on most realizations of these networks belongs to the I, ID, or MD class for $k=2,\,4,$ and 6 while the DI, IDI and DI2 classes are rare. It is also interesting to note that while the highest efficiency is observed for $k=7$ networks, the ratio of realizations that display ENAQT is the lowest for this particular type. The proportion of networks with noise-enhanced transport efficiency is found to depend on the topology in such a way that more (less) than half of 16-node Watts-Strogatz networks for $k<7$ ($k=7$) display some type of noise-enhanced transport. Moreover, as can be intuitively expected, noise-degraded efficiency is quite common among all realizations; the efficiency of almost half of $k=7$  and 1/4 to 1/3 of the realizations for the $k=2,4,6$ networks shows monotonous decay with increasing noise strength. 

In addition to the above findings, we also investigated whether any correlation could be established between 
efficiency $\eta$ and the different centrality and clustering measures of the networks.
This is motivated by the often asked question of whether the efficiency of transport -- and its noise dependence on graphs -- are correlated with the network structure described, e.g., by connectivity, regularity, and
various centrality measures. Studies in the perfect state transfer context indicate that network connectivity measures are not good indicators of the fidelity of state transfer~\cite{Razzoli2021,Meyer2015, Janmark2014}. However, although we calculated a large number of centrality measures and 3 different clustering coefficients for the networks we study in the current paper, we could not find any statistically significant relation with the efficiency.

\section{Conclusions}
\label{sec:conc}
In this work, we have presented an in-depth study of the topology dependence of excitation transfer efficiency in a multisite spin-boson system including both external noise and quantum bath. By solving the adopted variational polaron master equation, we have studied the transport efficiency in two types of complex networks -- in the random removal networks, which are constructed by deleting random edges from the completely connected network, and in the Watts-Strogatz small-world networks. We observe that the maximum efficiency attainable in both network classes is obtained when the network topology is modified in a small and specific way, i.e., severing the link between the source and the sink sites in the CCN and adding all nearest-neighbor interactions up to one layer less than complete connectedness for the Watts-Strogatz network. No manipulation of external noise driving or quantum environment was found to have an equally strong impact on efficiency in the networks studied. Our results indicate that the structure of the network is more important than the environmental conditions to achieve high transport efficiency. The main finding of our study might illustrate a possible mechanism that explains why highly efficient biological charge transfer complexes have a certain connectivity structure. Finally, from the efficiency distributions as a function of noise strength, we have observed that the noise dependence of transport efficiency in the studied Watts-Strogatz networks could be broadly classified into six different categories that show various modalities, from monotonous decay with increasing noise to ENAQT behavior. We believe that our findings may help to achieve higher transport efficiency mechanisms by changing network topology -- rather than engineering environmental models -- when increasing the size and complexity of networks within the quantum transport framework. 

\ack
A.K. acknowledges support from the Scientific and Technological Research Council of Turkey (TUBITAK) Project no:1002-120F011. M.A.C.R. acknowledges financial support from the Academy of Finland via the Centre of Excellence program (Project No. 336810).

\section*{Appendix}
\label{sec:appendix}

\begin{figure}[!hbt]
\centering
 \begin{subfigure}{.27\textwidth}
 \includegraphics[width=\linewidth]{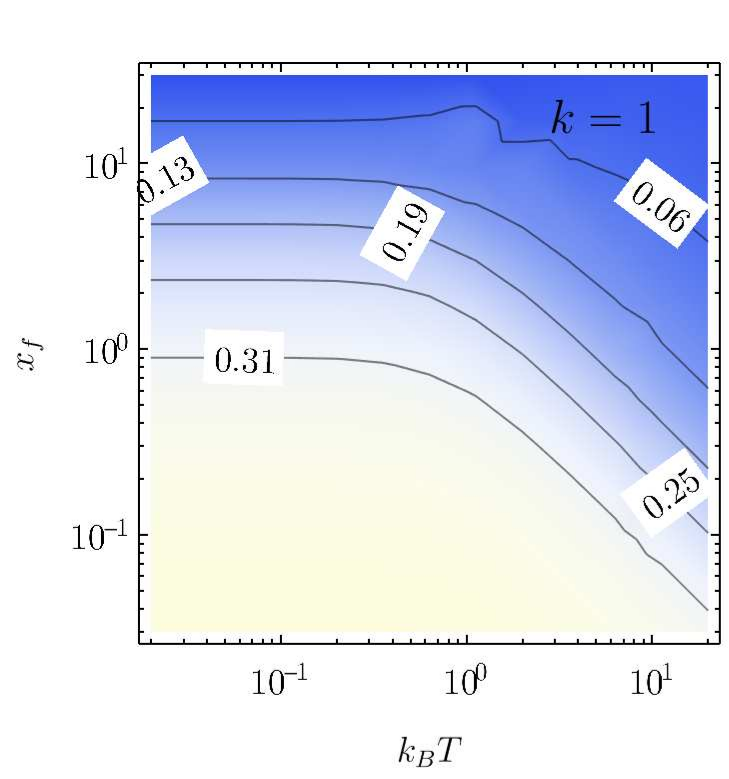}
 \caption{k=1} 
 \end{subfigure}
\centering
\begin{subfigure}{.27\textwidth}
 \includegraphics[width=\linewidth]{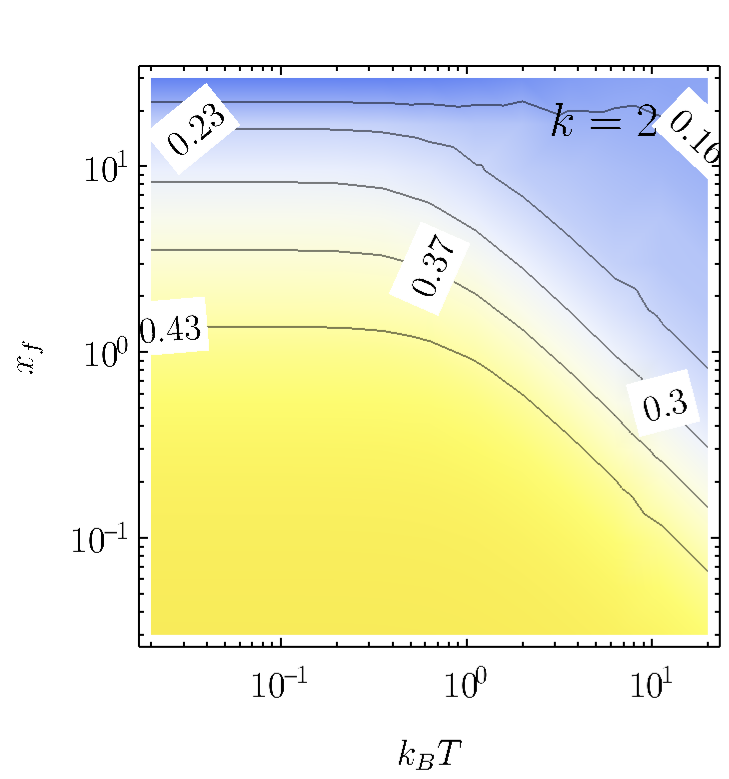}
 \caption{k=2}
\end{subfigure}
\centering
 \begin{subfigure}{.27\textwidth}
 \includegraphics[width=\linewidth]{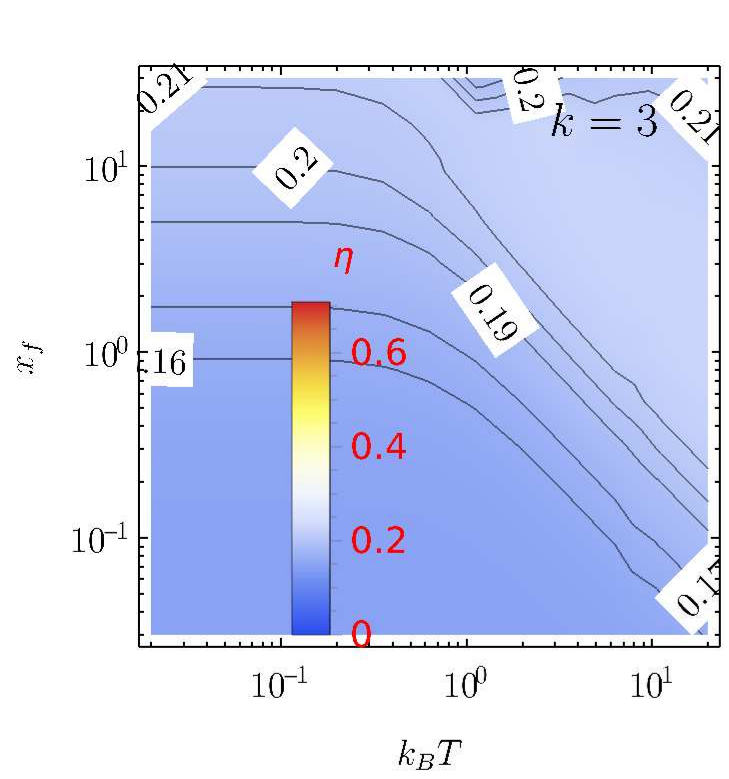}
 \caption{k=3} 
 \end{subfigure}
\centering
\begin{subfigure}{.27\textwidth}
 \includegraphics[width=\linewidth]{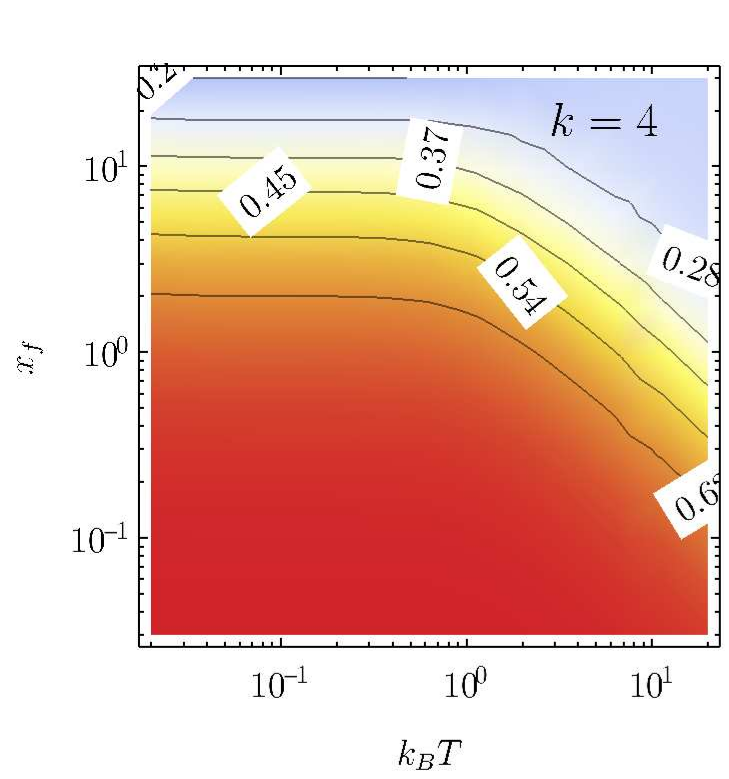}
 \caption{k=4}
\end{subfigure}
\centering
 \begin{subfigure}{.27\textwidth}
 \includegraphics[width=\linewidth]{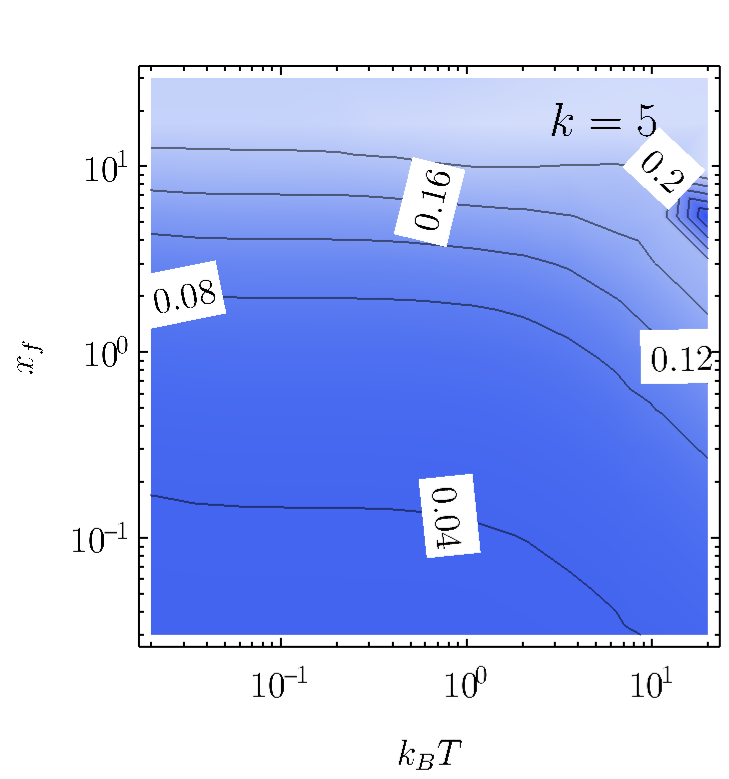}
 \caption{k=5} 
 \end{subfigure}
 \caption{Efficiency as function of environmental temperature $k_B T$ and the system-bath coupling coefficient $x_f$ for bath-only model for $V=2$. The same color scheme is used in all plots.}
 \label{fig:bath-kbt-xf}
\end{figure}

In this appendix, we examine the impact of the strength of the system-bath interaction and temperature on transport efficiency in WS networks. As the variational polaron approach allows one to treat all interaction strength regimes in one consistent formulation, and there is a large number of reported research on the weak- and strong-coupling regimes, the current study focuses on the intermediate regime. In this appendix, Fig.~\ref{fig:bath-kbt-xf} shows the efficiency $\eta$ as a function of $k_{B} T$ and the system-bath coupling coefficient $x_f$ for the bath-only model. The figure demonstrates that efficiency is more significantly affected by the parameter $k$ than by either the system-bath coupling at constant temperature or the temperature at constant bath coupling. Note that all plots in the figure use the same temperature map. Comparison of $k=3$ (mainly light blue, indicating low efficiency) with $k=4$ (mainly red, indicating high efficiency) highlights the significant change in efficiency with $k$.

\begin{figure}[!h]\centering
\begin{subfigure}{0.4\textwidth}
\includegraphics[width=\linewidth]{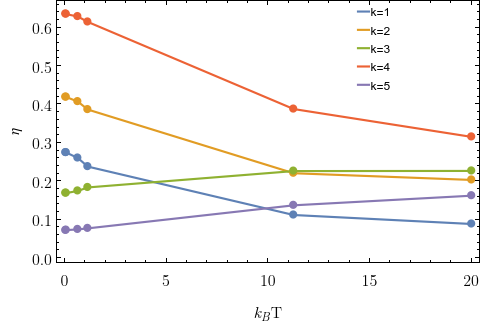}
\caption{$x_f=1$}
\end{subfigure}
\begin{subfigure}{0.4\textwidth}
\includegraphics[width=\linewidth]{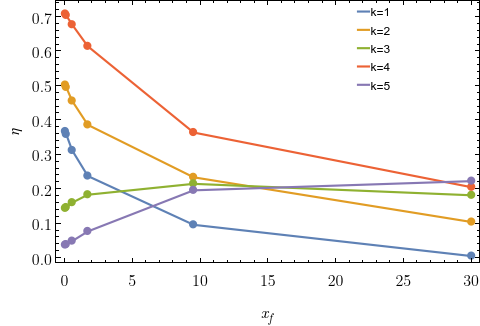}
\caption{$k_B T=1$}
\end{subfigure}
\caption{Transport efficiency as function of bath temperature $k_B T$ at a given $x_f$ (a) and system-environment coupling coefficient $x_f$ at a given $k_B T$ (b) at different $k$'s. }
\label{fig:eff3d}
\end{figure}

We present the $k$ dependence of the transport efficiency at constant system-bath interaction (Fig.~\ref{fig:eff3d}a) as function of bath temperature and at constant temperature (Fig.~\ref{fig:eff3d}b) as function of the system-bath coupling coefficient. Both plots indicate that maximum of $\eta$ would be obtained for $k=4$.

\providecommand{\newblock}{}

\end{document}